\documentclass[pra,twocolumn,showpacs,preprintnumbers,superscriptaddress]
{revtex4}

\usepackage{times}
\usepackage{bm}
\usepackage{float}
\usepackage{graphicx}
\usepackage{amsbsy}
\usepackage{amsmath,mathtools}
\usepackage{amsfonts}
\usepackage{amsthm}
\usepackage{xcolor}
\begin{document}
\theoremstyle{plain}
\newtheorem{theorem}{Theorem}
\newtheorem{lemma}[theorem]{Lemma}
\newtheorem{corollary}[theorem]{Corollary}
\newtheorem{proposition}[theorem]{Proposition}
\newtheorem{conjecture}[theorem]{Conjecture}
\theoremstyle{definition}
\newtheorem{definition}[theorem]{Definition}
\theoremstyle{remark}
\newtheorem*{remark}{Remark}
\newtheorem{example}{Example}

\title{Lower Bound of $l_{1}$ Norm of Coherence of Bipartite Qubit-Qudit System and its Application in the Detection of Entangled Tripartite Qudit-Qubit-Qudit System}
\author{Palash Garhwal, Pranav Chandhok, Satyabrata Adhikari}
\email{palashgarhwal402@gmail.com,pranav.chandhok2000@gmail.com,satyabrata@dtu.ac.in}
\affiliation{Delhi Technological University, Delhi-110042, Delhi, India}
\begin{abstract}
Quantum coherence and quantum entanglement are two strong pillars in quantum information theory. We study here for the possibility of any connection between these two important aspects of quantum mechanics while studying the entanglement detection problem for the detection of bipartite higher dimensional entangled states and multipartite entangled states. To achieve our goal, we derive the lower bound $L$ of $l_{1}$ norm of coherence of bipartite qubit-qudit system using the criterion that detect entanglement. Furthermore, we deduce the upper bound $U$ of $l_{1}$ norm of coherence of separable bipartite qubit-qudit system using the separability criterion. Thus, we find that if any $l_{1}$ norm of coherence of bipartite qubit-qudit system is greater than the upper bound $U$ then the given qubit-qudit state is entangled. Finally, we obtained the upper bound $U_{1}$ of $l_{1}$ norm of coherence of separable tripartite state lies either in $2 \otimes d \otimes d$ or $d \otimes 2 \otimes d$ or $d \otimes d \otimes 2$ dimensional Hilbert space using the upper bound $U$. We have shown that if the $l_{1}$ norm of coherence of any tripartite $qubit-qudit-qudit$ or $qudit-qubit-qudit$ or $qudit-qudit-qubit$ system is greater than the derived upper bound $U_{1}$ then the given tripartite system represent an entangled state.
%keywords: l1 norm of coherence, frobenius norm, Gell Mann  generalization, Holder's Inequality
\end{abstract}
\maketitle

\section{introduction}
Quantum coherence, which represent a superposition, is a very fundamental feature of quantum mechanics. On the other hand, quantum entanglement \cite{horodecki1} also is a very important and peculiar feature of quantum mechanics. It is peculiar in the sense that it possesses non-classical correlation \cite{piani} which cannot be simulated classically. Both of these features can be considered as a heart of quantum information theory \cite{nielsen} and quantum computing \cite{ladd}. The quantum coherence may originate to understand the behavior of the optical fields but now-a-days, the resource theory of coherence \cite{streltsov,winter}  has been developed to study the fundamental properties of quantum systems. Unlike entanglement, quantum coherence may exist in a single qubit also. Apart from quantum information theory, we can find the various application of quantum coherence in many other fields such as  Quantum thermodynamics \cite{brandao}, Quantum algorithms \cite{ekert}, Quantum metrology \cite{giorda}, Quantum biology \cite{huelga,engel}. Thus, by seeing the vast scope of quantum coherence in different areas, we need to understand few fundamental aspects of this important phenomenon such as (i) detection of quantum coherence \cite{yu,xzhang,nie} (ii) quantification of quantum coherence \cite{streltsov1,aberg,yuan,zhang} (iii) non-local advantage of coherence in performing quantum information processing task \cite{mondal,ding} and (iv) classification of different classes of entanglement in multipartite system using quantum coherence \cite{kairon,anu}. The recent development in the experiment for the detection and quantification of quantum coherence has been reviewed in \cite{wu}.\\
In \cite{zhu}, H. Zhu et.al. established one-one correspondence between coherence measures and entanglement measures and thus via this mapping they had obtained the experimentally achievable tight lower bound of generalised entanglement concurrence and coherence concurrence. In another work, A. Streltsov et.al. have introduced a measure of coherence for an arbitrary dimensional quantum system, which is expressed via maximum bipartite entanglement generated from incoherent operation performed on the system and incoherent ancilla \cite{streltsov2}. Thus, the interplay between coherence and entanglement under different operations in the resource theory have been studied in \cite{yamasaki}.\\
Entanglement detection problem considered to be as one of the crucial problem in quantum information theory and thus in this work, we use $l_{1}$ norm of coherence to study this problem. We derive few inequalities in terms of $l_{1}$ norm of coherence that may help to detect bipartite entanglement in qubit-qudit system. We then study the entanglement criterion in terms of $l_{1}$ norm of coherence for the tripartite state live in either $2 \otimes d \otimes d$ or $d \otimes 2 \otimes d$ or $d \otimes d \otimes 2$ dimensional Hilbert space. Since coherence depends on the basis of the Hilbert space where the quantum states are lying so we should stress here the fact that we have obtained all results in computational basis.\\
This work is distributed in different sections in the following way: In section-II, we have provided the earlier obtained results in the literature, which will be used in the latter sections. In section-III, we obtain the lower bound of $l_{1}$ norm of coherence of qubit-qubit system and then generalize this bound to qubit-qudit system. These bounds of $l_{1}$ norm of coherence may serve as an entanglement detection criterion. In section-IV, we obtain the upper bound of $l_{1}$ norm of coherence and then we show that it gives the necessary condition for the separability of bipartite qubit-qudit system. In section-V, we apply the obtained criterion in the previous section for the detection of the entangled tripartite state.
\section{Preliminaries: A Few Results}
\textbf{Result-1 \cite{bertlmann}:} The generalised Gell-Mann matrices (GGM) are the generalisation of Pauli matrices (in qubit system) and Gell-Mann matrices (in qutrit system). It can be used as a basis in higher dimensional system such as qudit system. The GGM matrices for qudit system can be expressed in terms of bra-ket vectors as
\begin{eqnarray}
\Lambda_{s}^{jk}= |j \rangle \langle k| + |k \rangle \langle j|,~~   1\leq j < k \leq d
\label{gmm1}
\end{eqnarray}
\begin{eqnarray}
\Lambda_{a}^{jk}= -i |j \rangle \langle k| + i |k \rangle \langle j|,~~   1\leq j < k \leq d
\label{gmm2}
\end{eqnarray}
\begin{eqnarray}
\Lambda^{l}=\frac{2}{l(l+1)}(\sum_{j=1}^{l} |j\rangle \langle j|-l|l+1\rangle\langle l+1|),~~ 1\leq l \leq d-1
\label{gmm3}
\end{eqnarray}
Here, $\Lambda_{s}^{jk}$,$\Lambda_{a}^{jk}$ and $\Lambda^{l}$ denoting the symmetric, asymmetric and diagonal GGM respectively. Symmetric and Asymmetric GGM are $\frac{d(d-1)}{2}$ in number while diagonal GGM are $d-1$ in number.\\
\textbf{Result-2:} If we choose $p,q > 1$ be any two numbers such that $\frac{1}{p}+\frac{1}{q} = 1$ holds then
\begin{eqnarray}
\sum_{i=1}^{n}|x_{i}y_{i}|\leq (\sum_{i=1}^{n}|x_{i}|^{p})^{\frac{1}{p}}(\sum_{i=1}^{n}|y_{i}|^{q})^{\frac{1}{q}}
\label{holdineq}
\end{eqnarray}
This inequality is known as Holder's inequality and it holds for all $(x_{1},x_{2},.....,x_{n}),(y_{1},y_{2},.....,y_{n})\in R^{n}~~or~~ C^{n}$.\\
For the case when $y_{i}=1,~~\forall i=1,2,....,n$, the inequality (\ref{holdineq}) reduces to
\begin{eqnarray}
\sum_{i=1}^{n}|x_{i}|\leq n^{\frac{1}{q}}(\sum_{i=1}^{n}|x_{i}|^{p})^{\frac{1}{p}}
\label{holdineq1}
\end{eqnarray}
Further, if $p=q=2$, then the inequality (\ref{holdineq1}) can be expressed as
\begin{eqnarray}
\sum_{i=1}^{n}|x_{i}|\leq n^{\frac{1}{2}}(\sum_{i=1}^{n}|x_{i}|^{2})^{\frac{1}{2}}
\label{holdineq2}
\end{eqnarray}
\textbf{Result-3 \cite{adhikari}:} Assume a bipartite state $\rho_{AB}$, which can be expressed as $\rho_{AB} =
\begin{bmatrix}
X & Y\\
Y^{\dagger} & Z\\
\end{bmatrix}$, where $X\geq 0$, $Y$ and $Z\geq 0$ denote the matrices of order $d$. If the block matrix satisfies the inequality
\begin{eqnarray}
Tr(XZ)< ||Y||_{2}^{2}
\label{adhikari}
\end{eqnarray}
then the state $\rho_{AB}$ is an entangled state in $2\otimes d$ dimensional Hilbert space.\\
\textbf{Result-4 \cite{johnston}:} Let $U$, $V$, $W$ denote the matrices of order $d$ such that $U$ and $W$ are positive semi-definite matrices. If
the block matrix $\varrho_{AB} =
\begin{bmatrix}
U & V\\
V^{\dagger} & W\\
\end{bmatrix}$ represent a $2\otimes d$ dimensional separable state, then
\begin{eqnarray}
||V||_{2}^{2}< \lambda_{min}(U)\lambda_{min}(W)
\label{johnston}
\end{eqnarray}
holds.

\section{Entanglement criteria for qubit-qubit and qubit-qudit system in terms of $l_{1}$ norm of coherence}
\noindent To quantify the coherence, it is desirable to first specify the conditions that should be satisfied by a proper measure of coherence. In this direction, Baumgratz et al. \cite{baumgratz} proposed that any proper measure of the coherence $C$ must satisfy the following conditions:\\
1. $C(\rho) \geq 0$ for all quantum states $\rho$ and $C(\rho) = 0$ if and only if $\rho \in I$, where $I$ denoting the set of incoherent states.\\
2. $C$ must be monotonic under incoherent completely positive and trace preserving maps.\\
3. $C$ must be monotonic for average coherence under subselection based on measurements outcomes.\\
4. $C$ must be convex.\\
$l_{1}$ norm of coherence is a very popular measure of coherence and it satisfies all the above conditions. Thus it serve as a good quantifier of coherence and may be defined as the summation of modulus of the off-diagonal terms of given quantum state. Mathematically, if a quantum state described by the density operator $\rho$ then $l_{1}$ norm of coherence of $\rho$ is defined as
\begin{eqnarray}
C_{l_{1}}(\rho) = \sum_{i\neq j} |\rho_{ij}|
\end{eqnarray}
\subsection{Qubit-Qubit System}
Let us consider a two-qubit bipartite quantum state $\rho^{2\otimes 2}$ expressed as
\begin{equation}
\rho^{2\otimes 2} =
\begin{bmatrix}
A & D\\
D^{\dagger} & B\\
\end{bmatrix}
\end{equation}
where the matrices $A$, $B$ and $D$ are given by
\begin{equation}
A =
\begin{pmatrix}
\rho_{11} & \rho_{12}\\
\rho_{12}^{\dagger} & \rho_{22}\\
\end{pmatrix},B =
\begin{pmatrix}
\rho_{33} & \rho_{34}\\
\rho_{34}^{\dagger} & \rho_{44}\\
\end{pmatrix},D =
\begin{pmatrix}
\rho_{13} & \rho_{14}\\
\rho_{23} & \rho_{24}\\
\end{pmatrix}
\end{equation}
The $l_{1}$ norm of coherence of the state $\rho^{2\otimes 2}$ in the computational basis is given by
\begin{eqnarray}
C_{l_{1}}(\rho^{2\otimes 2})=2(|\rho_{12}|+|\rho_{13}|+|\rho_{14}|+|\rho_{23}|+|\rho_{24}|+|\rho_{34}|)
\label{coh21}
\end{eqnarray}
Since $0\leq \rho_{ij}\leq 1,~~ i<j~~ \textrm{and} ~~\forall i,j=1,2,3,4$ so equation (\ref{coh21}) reduces to the inequality as
\begin{eqnarray}
C_{l_{1}}(\rho^{2\otimes 2})&\geq& 2(|\rho_{12}|+|\rho_{34}|+|\rho_{13}|^{2}+|\rho_{23}|^{2}+|\rho_{14}|^{2}\nonumber\\&+&|\rho_{24}|^{2})\nonumber\\&=&
2(|\rho_{12}|+|\rho_{34}|)+2 Tr(D^{\dagger}D)\nonumber\\&=& 2(|\rho_{12}|+|\rho_{34}|)+2||D||_{2}^{2}
\label{coh22}
\end{eqnarray}
where $||D||_{2}^{2}$ denote the Frobenius norm of $D$ and defined as $||D||_{2}^{2}=Tr(D^{\dagger}D)$.\\
From Result-3 given in (\ref{adhikari}), we know that the state $\rho^{2\otimes 2}$ is entangled if
\begin{eqnarray}
||D||_{2}^{2}> Tr(AB)
\label{entcond1}
\end{eqnarray}
Therefore, using the entanglement criterion (\ref{entcond1}), the inequality (\ref{coh22}) reduces to
\begin{eqnarray}
C_{l_{1}}(\rho^{2\otimes 2})&>& 2(|\rho_{12}|+|\rho_{34}|)+2Tr(AB)
\label{coh23}
\end{eqnarray}
Thus, if the inequality (\ref{coh23}) is satisfied by any arbitrary bipartite two-qubit state then that state must be an entangled state.
The inequality (\ref{coh23}) can further be re-expressed in terms of the matrices $A$ and $B$ as
\begin{eqnarray}
C_{l_{1}}(\rho^{2\otimes 2})&>& 2(|\rho_{12}|+|\rho_{34}|)+2Tr(AB)\nonumber\\ &\geq& 2(Re(\rho_{12})+\rho_{34}))+2Tr(AB)\nonumber\\&=&
Tr(A\sigma_{x})+Tr(B\sigma_{x})+2Tr(AB)\nonumber\\&=& Tr[(A+B)\sigma_{x}+2AB]
\label{coh24}
\end{eqnarray}
\textbf{Theorem-1:} If the $l_{1}$ norm of a coherence of a two-qubit state $\rho^{2\otimes 2}$ satisfies the inequality
\begin{eqnarray}
C_{l_{1}}(\rho^{2\otimes 2})&>& Tr[(A+B)\sigma_{x}+2AB]
\label{th1}
\end{eqnarray}
then the state $\rho^{2\otimes 2}$ is an entangled state.
\subsection{Qubit-Qudit System}
\noindent In this section, we generalize the result obtained in the previous section for Qubit-Qubit system. To start with the generalisation procedure, let us consider a bipartite quantum state in $2 \otimes d$ dimensional Hilbert space described by the density operator $\rho^{2 \otimes d}$, which is given by
\begin{eqnarray}
\rho^{2 \otimes d} =
\begin{pmatrix}
\rho_{1,1} & ...... & \rho_{1,d} & \rho_{1,d+1} & ...... & \rho_{1,2d}\\
...... & ...... & ...... & ...... & ......  & ......\\
...... & ...... & ...... & ...... & ......  & ......\\
\rho_{1,d}^{*} & ...... & \rho_{d,d} & \rho_{d,d+1} & ...... & \rho_{d,2d}\\
\rho_{1,d+1}^{*} & ...... & \rho_{d,d+1}^{*} & \rho_{d+1,d+1} & ...... & \rho_{d+1,2d}\\
...... & ...... & ...... & ...... & ...... & ......\\
...... & ...... & ...... & ...... & ...... & ......\\
\rho_{1,2d}^{*} & ...... & \rho_{d,2d}^{*} & \rho_{d+1,2d}^{*} & ...... & \rho_{2d,2d}\\
\end{pmatrix}
\label{denmat1}
\end{eqnarray}
In block matrix notation, the state $\rho^{2 \otimes d}$ can be expressed as
\begin{eqnarray}
\rho^{2 \otimes d}=
\begin{bmatrix}
 P & Q\\
 Q^* & R\\
\end{bmatrix}
\end{eqnarray}
where $P$, $Q$, $R$ denote the matrices of order $d$ and they are given by
\begin{eqnarray}
&&P =
\begin{pmatrix}
\rho_{1,1} & ...... & \rho_{1,d}\\
...... & ...... & ...... \\
...... & ...... & ...... \\
\rho_{1,d}^{*} & ...... & \rho_{d,d}\\
\end{pmatrix},Q =
\begin{pmatrix}
\rho_{1,d+1} & ...... & \rho_{1,2d}\\
...... & ...... & ...... \\
...... & ...... & ...... \\
\rho_{d,d+1} & ...... & \rho_{d,2d}
\end{pmatrix},\nonumber\\&&R =
\begin{pmatrix}
\rho_{d+1,d+1} & ...... & \rho_{d+1,2d}\\
...... & ...... & ...... \\
...... & ...... & ...... \\
\rho_{d+1,2d}^{*} & ...... & \rho_{2d,2d}\\
\end{pmatrix}
\label{blockmat1}
\end{eqnarray}
The $l_{1}$ norm of coherence of the state $\rho^{2 \otimes d}$ in the computational basis is given by
\begin{eqnarray}
C_{l_{1}}(\rho^{2 \otimes d}) = \sum_{i \neq j} |\rho^{2 \otimes d}_{ij}|
\label{cohd1}
\end{eqnarray}
The off-diagonal elements appearing in (\ref{cohd1}) split into the off diagonal elements of the matrices $P$ and $R$ and all elements
of the matrices $Q$ and $Q^{\dagger}$. Thus, we have
\begin{eqnarray}
C_{l_{1}}(\rho^{2 \otimes d}) &=& \underbrace{\sum_{\substack{{i \neq j}\\ {i,j = 1}}}^{\substack{{d}}} |\rho^{2 \otimes d}_{ij}|}_{\substack{\text{Sum of off-diagonal}\\ \text{elements of P}}}+ \underbrace{\sum_{\substack{{i \neq j}\\ {i,j = d+1}}}^{\substack{{2d}}} |\rho^{2 \otimes d}_{ij}|}_{\substack{\text{Sum of off-diagonal}\\ \text{elements of R}}}\nonumber\\&+& 2\underbrace{\sum_{\substack{{i \neq j}\\ {i=1}}}^{\substack{{d}}}\sum_{\substack{{j=d+1}}}^{\substack{{2d}}} |\rho^{2 \otimes d}_{ij}|}_{\substack{\text{Sum of elements}\\ {of Q \& Q^{\dagger}}}}\nonumber\\&=&
2[\sum_{\substack{{i < j}\\ {i,j = 1}}}^{\substack{{d}}} |\rho^{2 \otimes d}_{ij}|+ \sum_{\substack{{i < j}\\ {i,j = d+1}}}^{\substack{{2d}}} |\rho^{2 \otimes d}_{ij}|\nonumber\\&+& \sum_{i=1}^{d}\sum_{j=d+1}^{2d} |\rho^{2 \otimes d}_{ij}|]
\label{cohd2}
\end{eqnarray}
Since $|\rho^{2 \otimes d}_{ij}| \in [0,1],~~\forall i,j=1,....2d$, so the inequality (\ref{cohd2}) takes the form
\begin{eqnarray}
C_{l_{1}}(\rho^{2 \otimes d}) &\geq& 2[\sum_{\substack{{i < j}\\ {i,j = 1}}}^{\substack{{d}}} |\rho^{2 \otimes d}_{ij}|+ \sum_{\substack{{i < j}\\ {i,j = d+1}}}^{\substack{{2d}}} |\rho^{2 \otimes d}_{ij}|] + 2Tr(Q^{\dagger}Q)\nonumber\\ &=& 2[\sum_{\substack{{i < j}\\ {i,j = 1}}}^{\substack{{d}}} |\rho^{2 \otimes d}_{ij}|+ \sum_{\substack{{i < j}\\ {i,j = d+1}}}^{\substack{{2d}}} |\rho^{2 \otimes d}_{ij}|] + 2||Q||_2^2\nonumber\\ &>& 2[\sum_{\substack{{i < j}\\ {i,j = 1}}}^{\substack{{d}}} |\rho^{2 \otimes d}_{ij}|+ \sum_{\substack{{i < j}\\ {i,j = d+1}}}^{\substack{{2d}}} |\rho^{2 \otimes d}_{ij}|]\nonumber\\&+& 2Tr(PR)
\label{cohd3}
\end{eqnarray}
The last inequality follows from the entanglement criterion given in (\ref{adhikari}). The inequality (\ref{cohd3}) may be expressed in terms of the generalised Gell-Mann symmetric matrix as
\begin{eqnarray}
C_{l_{1}}(\rho^{2 \otimes d}) &\geq& 2[\sum_{\substack{{i < j}\\ {i,j = 1}}}^{\substack{{d}}} Re(\rho^{2 \otimes d}_{ij})+ \sum_{\substack{{i < j}\\ {i,j = d+1}}}^{\substack{{2d}}} Re(\rho^{2 \otimes d}_{ij})] + 2Tr(PR)\nonumber\\&=&Tr(P\sum_{\substack{{j < k}\\ {j,k = 1}}}^{{\substack{{d}}}} \Lambda_s^{jk})+Tr(R\sum_{\substack{{j < k}\\ {j,k = 1}}}^{\substack{{d}}} \Lambda_s^{jk}) + 2Tr(PR)\nonumber\\&=&Tr[(P+R)(\sum_{\substack{{j < k}\\ {j,k = 1}}}^{{\substack{{d}}}} \Lambda_s^{jk})]+2Tr[PR]
\end{eqnarray}
where $\Lambda_s^{jk}$ denote the $\frac{d(d-1)}{2}$ symmetric generalised Gell-Mann matrices given in (\ref{gmm1}).\\
\textbf{Theorem-2:} If an arbitrary qubit-qudit state $\rho^{2\otimes d}$ lying in $2\otimes d$ dimensional Hilbert space and its $l_{1}$ norm of coherence satisfies the inequality
\begin{eqnarray}
C_{l_{1}}(\rho^{2 \otimes d}) &\geq& Tr[(P+R)(\sum_{\substack{{j < k}\\ {j,k = 1}}}^{{\substack{{d}}}} \Lambda_s^{jk})]+2Tr[PR]
\label{th2}
\end{eqnarray}
then the state described by the density operator $\rho^{2\otimes d}$ is an entangled state.
\subsection{Example}
\textbf{Example-1:} Let us consider a bipartite $2 \otimes 2$ system described by the density operator $\chi_{1}^{2\otimes 2}$. It is given by
\begin{eqnarray}
\chi_{1}^{2\otimes 2}=\begin{pmatrix}
a & 0 & 0 & f\\
0 & b & c & 0\\
0 & c^{*} & d & 0\\
f^{*} & 0 & 0 & e\\
\end{pmatrix}, ~~ a+b+d+e=1
\end{eqnarray}
$\chi_{1}^{2\otimes 2}$ represent a quantum state if $bd \geq |c|^{2}$ and $ae \geq |f|^{2}$ \cite{mazhar}. The $l_{1}$ norm of coherence of $\chi_{1}^{2\otimes 2}$ is given by
\begin{eqnarray}
C_{l_{1}}(\chi_{1}^{2\otimes 2}) = 2(|c|+|f|)
\label{exl3}
\end{eqnarray}
Our task is now to calculate the R.H.S of the inequality given in (\ref{th1}). Therefore, we have
\begin{eqnarray}
R.H.S= Tr[(A+B)\sigma_{x}+2AB]=2 (ad+be)
\label{exl4}
\end{eqnarray}
Using Theorem-1, we can say that the state $\chi_{1}^{2\otimes 2}$ is entangled if
\begin{eqnarray}
 |c|+|f|>ad+be
 \label{exl5}
\end{eqnarray}
\textbf{Example-2:}Let us consider a quantum state $\chi_{2}^{2\otimes 4}$ in $2\otimes 4$ dimensional Hilbert space which is given by
\begin{eqnarray}
\chi_{2}^{2\otimes 4}=\begin{pmatrix}
 M & O\\
 O^{\dagger} & N\\
\end{pmatrix}
\end{eqnarray}
where the matrices $M$, $N$, $O$ are given by
\begin{eqnarray}
&&M=\begin{pmatrix}
\frac{a}{6a+1} & 0 & 0 & 0\\
0 & \frac{a}{6a+1} & 0 & 0\\
0 & 0 & \frac{a}{6a+1} & 0\\
0 & 0 & 0 & 0\\
\end{pmatrix},N=\begin{pmatrix}
0 & 0 & 0 & 0\\
0 & \frac{a}{6a+1} & 0 & 0\\
0 & 0 & \frac{a}{6a+1} & 0\\
0 & 0 & 0 & \frac{a+1}{6a+1}\\
\end{pmatrix},\nonumber\\&&O=\begin{pmatrix}
0 & 0 & 0 & \frac{a}{6a+1}\\
0 & 0 & \frac{a}{6a+1} & 0\\
0 & \frac{a}{6a+1} & 0 & 0\\
0 & 0 & 0 & 0\\
\end{pmatrix},~~ 0\leq a \leq 1
\end{eqnarray}
The state $\chi_{2}^{2\otimes 4}$ is shown to be an entangled state in \cite{adhikari}. We will now verify it again to justify our result, which is given in (\ref{th2}). Let us start with the $l_{1}$ norm of coherence of $\chi_{2}^{2\otimes 4}$, which is given by
\begin{eqnarray}
C_{l_{1}}(\chi_{2}^{2\otimes 4}) = \frac{6a}{6a+1}
\label{ex11}
\end{eqnarray}
The R.H.S of the inequality (\ref{th2}) can be calculated as
\begin{eqnarray}
Tr[(M+N)(\sum_{\substack{{j < k}\\ {j,k = 1}}}^{{\substack{{4}}}} \Lambda_s^{jk})]+2Tr[MN]=\frac{4a^{2}}{(6a+1)^{2}}
\label{ex12}
\end{eqnarray}
It can be easily verified that the inequality (\ref{th2}) is satisfied for the state $\chi_{2}^{2\otimes 4}$ when $0\leq a \leq 1$. Thus the state $C_{l_{1}}(\chi_{2}^{2\otimes 4})$ is proved to be an entangled state for $0\leq a \leq 1$.
\section{Separability criterion for a qubit-qudit state in terms of $l_{1}$ norm of coherence}
In this section, we derive the upper bound of the $l_{1}$ norm of coherence of a bipartite separable qubit-qudit state. In the derivation, we use the separability criterion given in (\ref{johnston}), which help us to derive the new separability condition in terms of $l_{1}$ norm of coherence.\\
Let us recall a qubit-qudit state described by the density operator $\rho^{2 \otimes d}$ given in (\ref{denmat1}) and assume it to be a separable state. The $l_{1}$ norm of coherence $C_{l_{1}}(\rho^{2 \otimes d})$ can be re-written as
\begin{eqnarray}
C_{l_{1}}(\rho^{2 \otimes d}) &=& \sum_{i \neq j} |\rho^{2 \otimes d}_{ij}|\nonumber\\&=& 2\sum_{\substack{{i < j}\\ {i,j = 1}}}^{\substack{{2d}}} |\rho^{2 \otimes d}_{ij}|\nonumber\\&+& 2\sum_{i=1}^{d}\sum_{j=d+1}^{2d} |\rho^{2 \otimes d}_{ij}|
\label{norm1}
\end{eqnarray}
Now, using the particular form (\ref{holdineq2}) of the Holder's inequality, the expression of $C_{l_{1}}(\rho^{2 \otimes d})$ given in (\ref{norm1}) become an inequality, which may be expressed as
\begin{eqnarray}
C_{l_{1}}(\rho^{2 \otimes d)}&\leq& 2\sqrt{\frac{d(d-1)}{2}}[(\sum_{\substack{{i < j}\\ {i,j = 1}}}^{\substack{{2d}}} |\rho^{2 \otimes d}_{ij}|^{2})^{\frac{1}{2}}+ \nonumber\\&&(\sum_{i=1}^{d}\sum_{j=d+1}^{2d} |\rho^{2 \otimes d}_{ij}|^{2})^{\frac{1}{2}}]
\label{cohd4}
\end{eqnarray}
Now, recalling the matrices $P$, $Q$ and $R$ from (\ref{blockmat1}) and then calculate $Tr(PP^{\dagger})$, $Tr(QQ^{\dagger})$
and $Tr(RR^{\dagger})$. The values of the traces are given by
\begin{eqnarray}
&&\|P\|_{2}^{2}=Tr(PP^\dagger)=\sum^{d}_{i=1} |\rho^{2 \otimes d}_{ii}|^2 + 2 \sum_{\substack{{i < j}\\ {i,j = 1}}}^{\substack{{d}}}|\rho^{2 \otimes d}_{ij}|^2\nonumber\\&&
\|R\|_{2}^{2}=Tr(RR^\dagger)=\sum^{2d}_{i=d+1} |\rho^{2 \otimes d}_{ii}|^2 + 2 \sum_{\substack{{i < j}\\ {i,j = d+1}}}^{\substack{{2d}}}|\rho^{2 \otimes d}_{ij}|^2 \nonumber\\&&
\|Q\|_{2}^{2}=Tr(QQ^\dagger)=\sum_{i=1}^{d}\sum_{j=d+1}^{2d} |\rho^{2 \otimes d}_{ij}|^2
\label{traces1}
\end{eqnarray}
Using (\ref{traces1}), the inequality (\ref{cohd4}) reduces to
\begin{eqnarray}
C_{l_{1}}(\rho^{2\otimes d}) &\leq& \sqrt{2d(d-1)}  [ (\|P\|_{2}^{2}+\|R\|_{2}^{2}-\sum^{2d}_{i=1} |\rho^{2 \otimes d}_{ii}|^2)^\frac{1}{2} \nonumber\\&& + ||Q||_{2}]
\label{cohd5}
\end{eqnarray}
If we impose the condition of separability (\ref{johnston}), then the inequality (\ref{cohd5}) further modified to
\begin{eqnarray}
C_{l_{1}}(\rho^{2\otimes d}) &\leq& \sqrt{2d(d-1)} [(\|P\|_{2}^{2}+\|R\|_{2}^{2}-\sum^{2d}_{i=1} |\rho^{2 \otimes d}_{ii}|^2)^\frac{1}{2} \nonumber\\&+& \sqrt{\lambda_{min}(P) \lambda_{min}(R)}]
\label{cohd6}
\end{eqnarray}
Thus, we are now in a position to state the following theorem:\\
\textbf{Theorem-3:} If any arbitrary qubit-qudit state described by the density operator $\rho^{2\otimes d}$ is separable then its $l_{1}$ norm of coherence satisfies the inequality
\begin{eqnarray}
C_{l_{1}}(\rho^{2\otimes d}) &\leq& \sqrt{2d(d-1)} [(\|P\|_{2}^{2}+\|R\|_{2}^{2}-\sum^{2d}_{i=1} |\rho^{2 \otimes d}_{ii}|^2)^\frac{1}{2} \nonumber\\&+& \sqrt{\lambda_{min}(P) \lambda_{min}(R)}]
\label{th3}
\end{eqnarray}
\textbf{Corollary-1:} If the inequality (\ref{th3}) is violated by an arbitrary qubit-qudit state $\varrho^{2\otimes d}$ then the state $\varrho^{2\otimes d}$ represent an entangled state.
\section{Detection of tripartite entangled state: An application of the separability condition of the qubit-qudit system}
In this section, we obtain the upper bound of $l_{1}$ norm of coherence of separable tripartite state lies either in $2 \otimes d \otimes d$ or $d \otimes 2 \otimes d$ or $d \otimes d \otimes 2$ dimensional Hilbert space using the upper bound of $l_{1}$ norm of coherence of qubit-qudit system obtained in the previous section. We further show that if the $l_{1}$ norm of coherence of any tripartite $qubit-qudit-qudit$ or $qudit-qubit-qudit$ or $qudit-qudit-qubit$ system is greater than the derived upper bound then the given tripartite system represent an entangled state. We derive here the upper bound by considering the tripartite state lying in $d\otimes 2 \otimes d$ dimensional Hilbert space. The derivation will be similar for the tripartite state either lying in $2\otimes d \otimes d$ or $d\otimes d \otimes 2$ dimensional Hilbert space.
\subsection{Upper bound of the $l_{1}$ norm of coherence of the separable tripartite system}
To investigate this, we start with the tripartite separable $qudit-qubit-qudit$ system, which is expressed in the form as
\begin{eqnarray}
\varrho_{ABC}=\sum_{i}p_{i}\varrho^{(i)}_{A}\otimes \varrho^{(i)}_{BC},~~\sum_{i}p_{i}=1
\label{tripartite2}
\end{eqnarray}
where the single qudit-system is described by the density operator $\varrho^{(i)}_{A}$ and the qubit-qudit system is described by the density operator $\varrho^{(i)}_{BC}$. It should be noted here that qubit-qudit system $BC$ represent a separable state.\\
To achieve the separability condition, let us start with the $l_{1}$ norm of coherence of the separable state $\varrho_{ABC}$. It is given by
\begin{eqnarray}
C_{l_{1}}(\varrho_{ABC})&=& C_{l_{1}}(\sum_{i}p_{i}\varrho^{(i)}_{A}\otimes \varrho^{(i)}_{BC})\nonumber\\&\leq&
\sum_{i} C_{l_{1}}(p_{i}\varrho^{(i)}_{A}\otimes \varrho^{(i)}_{BC}))\nonumber\\&=& \sum_{i} p_{i}[C_{l_{1}}(\varrho^{(i)}_{A})+C_{l_{1}} (\varrho^{(i)}_{BC})(1+C_{l_{1}}(\varrho^{(i)}_{A}))]
\label{tripartite3}
\end{eqnarray}
The inequality in the second line follows from the convexity property of the $l_{1}$ norm of coherence \cite{baumgratz} and the last line follows from the result given in \cite{anu}.\\
Now, applying $l_{1}$ norm of coherence based separability criterion (\ref{th3}) on $\varrho^{(i)}_{BC}$ for each $i$, the inequality (\ref{tripartite3}) reduces to
\begin{eqnarray}
C_{l_{1}}(\varrho_{ABC})&\leq&\sum_{i} p_{i}[C_{l_{1}}(\varrho^{(i)}_{A})+\sqrt{2d(d-1)} [(\|P_{BC}^{(i)}\|_{2}^{2}\nonumber\\&+&\|R_{BC}^{(i)}\|_{2}^{2}-\sum^{2d}_{j=1} |(\varrho^{(i)}_{BC})_{jj}|^2)^\frac{1}{2} \nonumber\\&+& \sqrt{\lambda_{min}(P_{BC}^{(i)}) \lambda_{min}(R_{BC}^{(i)})}]\times\nonumber\\&&(1+C_{l_{1}}(\varrho^{(i)}_{A}))]
\label{tripartite4}
\end{eqnarray}
where the qubit-qudit density matrix $\varrho^{(i)}_{BC}$ can be expressed in the block matrix form as
\begin{eqnarray}
\varrho^{(i)}_{BC}=
\begin{bmatrix}
 P^{(i)}_{BC} & Q^{(i)}_{BC}\\
 (Q^{(i)})_{BC}^* & R^{(i)}_{BC}\\
\end{bmatrix}
\end{eqnarray}
The matrices $P^{(i)}_{BC}$, $Q^{(i)}_{BC}$ and $R^{(i)}_{BC}$ can be given in the form (\ref{blockmat1}). It can be easily seen that the result given in (\ref{tripartite4}) will be true for the other bipartition such as $B-CA$ and $C-AB$ bipartition. The result (\ref{tripartite4}) for the bipartition $B-CA$ and $C-AB$ can be re-written as
\begin{eqnarray}
C_{l_{1}}(\varrho_{ABC})&\leq&\sum_{i} p_{i}[C_{l_{1}}(\varrho^{(i)}_{B})+\sqrt{2d(d-1)} [(\|P_{CA}^{(i)}\|_{2}^{2}\nonumber\\&+&\|R_{CA}^{(i)}\|_{2}^{2}-\sum^{2d}_{j=1} |(\varrho^{(i)}_{CA})_{jj}|^2)^\frac{1}{2} \nonumber\\&+& \sqrt{\lambda_{min}(P_{CA}^{(i)}) \lambda_{min}(R_{CA}^{(i)})}]\times\nonumber\\&&(1+C_{l_{1}}(\varrho^{(i)}_{B}))]
\label{tripartite5}
\end{eqnarray}
\begin{eqnarray}
C_{l_{1}}(\varrho_{ABC})&\leq&\sum_{i} p_{i}[C_{l_{1}}(\varrho^{(i)}_{C})+\sqrt{2d(d-1)} [(\|P_{AB}^{(i)}\|_{2}^{2}\nonumber\\&+&\|R_{AB}^{(i)}\|_{2}^{2}-\sum^{2d}_{j=1} |(\varrho^{(i)}_{AB})_{jj}|^2)^\frac{1}{2} \nonumber\\&+& \sqrt{\lambda_{min}(P_{AB}^{(i)}) \lambda_{min}(R_{AB}^{(i)})}]\times\nonumber\\&&(1+C_{l_{1}}(\varrho^{(i)}_{C}))]
\label{tripartite6}
\end{eqnarray}
Thus, if the tripartite system is separable then the separability condition in terms of $l_{1}$ nor of coherence may be stated as follows:\\
\textbf{Theorem-4:} If a tripartite separable state $\sigma_{ABC}$ lies in either $d \otimes 2 \otimes d$ or $d \otimes d \otimes 2$ or $2 \otimes d \otimes d$ dimensional Hilbert space then for $x\neq y\neq z,x,y,z=A,B,C$, the separable state $\sigma_{ABC}$ satisfies the following inequality
\begin{eqnarray}
C_{l_{1}}(\varrho_{ABC})&\leq&\sum_{i} p_{i}[C_{l_{1}}(\varrho^{(i)}_{x})+\sqrt{2d(d-1)} [(\|P_{yz}^{(i)}\|_{2}^{2}\nonumber\\&+&\|R_{yz}^{(i)}\|_{2}^{2}-\sum^{2d}_{j=1} |(\varrho^{(i)}_{yz})_{jj}|^2)^\frac{1}{2} \nonumber\\&+& \sqrt{\lambda_{min}(P_{yz}^{(i)}) \lambda_{min}(R_{yz}^{(i)})}]\times\nonumber\\&&(1+C_{l_{1}}(\varrho^{(i)}_{x}))]
\label{th4}
\end{eqnarray}
\textbf{Corollary-2:} If any tripartite state $\varsigma_{ABC}$ violate the inequality (\ref{th4}) then the given state $\varsigma_{ABC}$ is entangled.

\subsection{Illustration}
\textbf{Illustration-1:} Let us consider a three-qubit state $\xi_{ABC}^{(1)}$ given by
\begin{eqnarray}
\xi_{ABC}^{(1)} &=& p|0\rangle\langle 0| \otimes |\phi^+ \rangle\langle \phi^+| + (1-p)|1\rangle\langle 1| \otimes |\phi^- \rangle\langle \phi^-|
\label{3s1}
\end{eqnarray}
where $p\in [0,1]$, $|\phi^+\rangle = \frac{1}{\sqrt{2}}(|00\rangle + |11\rangle|)$ and $|\phi^-\rangle =\frac{1}{\sqrt{2}}(|00\rangle - |11\rangle|)$.\\
It can be easily verify that the $l_{1}$ norm of coherence of the single qubit state $|0\rangle\langle 0|$ and $|1\rangle\langle 1|$ are equal to zero.\\
In matrix notation, the state $\xi^{ABC}$ can be expressed as
\begin{eqnarray}
\xi_{ABC}^{(1)} =
\begin{pmatrix}
\frac{p}{2} & 0 & 0 & \frac{p}{2} & 0 & 0 & 0 & 0\\
0 & 0 & 0 & 0 & 0 & 0 & 0 & 0\\
0 & 0 & 0 & 0 & 0 & 0 & 0 & 0\\
\frac{p}{2} & 0 & 0 & \frac{p}{2} & 0 & 0 & 0 & 0\\
0 & 0 & 0 & 0 & \frac{1-p}{2} & 0 & 0 & \frac{-(1-p)}{2}\\
0 & 0 & 0 & 0 & 0 & 0 & 0 & 0\\
0 & 0 & 0 & 0 & 0 & 0 & 0 & 0\\
0 & 0 & 0 & 0 & \frac{-(1-p)}{2} & 0 & 0 & \frac{1-p}{2}\\
\end{pmatrix}\\
\end{eqnarray}
The $l_{1}$ norm of coherence of $\xi_{ABC}^{(1)}$ is given by
\begin{eqnarray}
C_{l_{1}}(\xi_{ABC}^{(1)}) = 1
\label{3s2}
\end{eqnarray}
The matrix representation of two qubit state $|\phi^+ \rangle\langle \phi^+|$ is given by
\begin{eqnarray}
|\phi^+ \rangle\langle \phi^+| &=&
\begin{pmatrix}
P_{BC}^{(1)} & Q_{BC}^{(1)}\\
(Q_{BC}^{(1)})^{\dagger} & R_{BC}^{(1)}
\end{pmatrix}\nonumber\\&\equiv&
\begin{pmatrix}
1/2 & 0 & 0 & 1/2\\
0 & 0 & 0 & 0\\
0 & 0 & 0 & 0\\
1/2 & 0 & 0 & 1/2\\
\end{pmatrix}
\end{eqnarray}
The minimum eigenvalue and Frobenius norm of the matrices $P_{BC}^{(1)}$ and $R_{BC}^{(1)}$ are given by
\begin{eqnarray}
&&\lambda_{min}(P_{BC}^{(1)})=\lambda_{min}(R_{BC}^{(1)})=0\nonumber\\&&
\|P_{BC}^{(1)}\|_{2}=\frac{1}{2}, \|R_{BC}^{(1)}\|_{2}=\frac{1}{2}
\label{3s3}
\end{eqnarray}
The matrix representation of two qubit state $|\phi^- \rangle\langle \phi^-|$ is given by
\begin{eqnarray}
|\phi^- \rangle\langle \phi^-| &=&
\begin{pmatrix}
P_{BC}^{(2)} & Q_{BC}^{(2)}\\
(Q_{BC}^{(2)})^{\dagger} & R_{BC}^{(2)}
\end{pmatrix}\nonumber\\&\equiv&
\begin{pmatrix}
1/2 & 0 & 0 & -1/2\\
0 & 0 & 0 & 0\\
0 & 0 & 0 & 0\\
-1/2 & 0 & 0 & 1/2\\
\end{pmatrix}
\end{eqnarray}
The minimum eigenvalue and Frobenius norm of the matrices $P_{BC}^{(2)}$ and $R_{BC}^{(2)}$ are given by
\begin{eqnarray}
&&\lambda_{min}(P_{BC}^{(2)})=\lambda_{min}(R_{BC}^{(2)})=0\nonumber\\&&
\|P_{BC}^{(2)}\|_{2}=\frac{1}{2}, \|R_{BC}^{(2)}\|_{2}=\frac{1}{2}
\label{3s4}
\end{eqnarray}
The R.H.S of the inequality (\ref{th4}) can be calculated using the values given in (\ref{3s3}) and (\ref{3s4}) and found out to be $0$ while from (\ref{3s2}),the L.H.S of the inequality (\ref{th4}) is found out to be $1$. Thus the inequality (\ref{th4}) is violated by $\xi^{ABC}$ and hence using the corollary-2, we can infer that the state $\xi^{ABC}$ is an entangled state.\\
\textbf{Illustration-2:} Let us consider a three-qubit state $\xi_{ABC}^{(2)}$ expressed in the form as
\begin{eqnarray}
\xi_{ABC}^{(2)} = p |\psi_{1}\rangle_{ABC}\langle \psi_{1}| + (1-p) |\psi_{2}\rangle_{ABC}\langle \psi_{2}|
\label{3ss1}
\end{eqnarray}
where $p\in [0,1]$, $|\psi_{1}\rangle_{ABC} = \frac{1}{\sqrt{5}}(|000\rangle + |100\rangle + |110\rangle + \sqrt{2}|111\rangle)$ and $|\psi_{2}\rangle_{ABC} = \frac{1}{\sqrt{5}}(|000\rangle + |100\rangle - |101\rangle + \sqrt{2}|110\rangle)$.\\
The $l_{1}$ norm of coherence of the three-qubit state $\xi_{ABC}^{(2)}$ is given by
\begin{eqnarray}
C_{l_{1}}(\xi_{ABC}^{(2)}) = \frac{6(1+\sqrt{2})}{5}
\label{3sslhs}
\end{eqnarray}
The $l_{1}$ norm of coherence of the single qubit states $\rho_{A}^{(1)}=Tr_{BC}(|\psi_{1}\rangle_{ABC}\langle \psi_{1}\rangle_{ABC}|)$, $\rho_{A}^{(2)}=Tr_{BC}(|\psi_{2}\rangle_{ABC}\langle \psi_{2}\rangle_{ABC}|)$ are given by
\begin{eqnarray}
C_{l_{1}}(\rho_{A}^{(1)}) = \frac{2}{5}, ~~ C_{l_{1}}(\rho_{A}^{(2)}) = \frac{2}{5}
\label{3ss2}
\end{eqnarray}
The reduced density operator $\rho_{BC}^{(1)}= Tr_{A}(|\psi_{1}\rangle_{ABC}\langle \psi_{1}\rangle_{ABC}|)$ is given by
\begin{eqnarray}
\rho_{BC}^{(1)} &=& \begin{pmatrix}
\frac{2}{5} & 0 & \frac{1}{5} & \frac{\sqrt{2}}{5}\\
0 & 0 & 0 & 0\\
\frac{1}{5} & 0 & \frac{1}{5} & \frac{\sqrt{2}}{5}\\
\frac{\sqrt{2}}{5} & 0 & \frac{\sqrt{2}}{5} & \frac{2}{5}
\end{pmatrix}\equiv \begin{pmatrix}
P_{BC}^{(1)} & Q_{BC}^{(1)}\\
(Q_{BC}^{(1)})^{\dagger} & R_{BC}^{(1)}
\end{pmatrix}
\label{3ss35}
\end{eqnarray}
The minimum eigenvalue and Frobenius norm of the matrices $P_{BC}^{(1)}$ and $R_{BC}^{(1)}$ are given by
\begin{eqnarray}
&&\lambda_{min}(P_{BC}^{(1)})=0, \lambda_{min}(R_{BC}^{(1)})= \nonumber\\&&
\|P_{BC}^{(1)}\|_{2}=\frac{4}{25}, \|R_{BC}^{(2)}\|_{2}=\frac{9}{25}
\label{3ss3}
\end{eqnarray}
The reduced density operator $\rho_{BC}^{(2)}= Tr_{A}(|\psi_{2}\rangle_{ABC}\langle \psi_{2}\rangle_{ABC}|)$ is given by
\begin{eqnarray}
\rho_{BC}^{(2)} &=& \begin{pmatrix}
\frac{2}{5} & \frac{-1}{5} & \frac{\sqrt{2}}{5} & 0\\
\frac{-1}{5} & \frac{1}{5} & \frac{-\sqrt{2}}{5} & 0\\
\frac{\sqrt{2}}{5} & \frac{-\sqrt{2}}{5} & \frac{2}{5} & 0\\
0 & 0 & 0 & 0
\end{pmatrix}\equiv \begin{pmatrix}
P_{BC}^{(2)} & Q_{BC}^{(2)}\\
(Q_{BC}^{(2)})^{\dagger} & R_{BC}^{(2)}
\end{pmatrix}
\label{3ss25}
\end{eqnarray}
The minimum eigenvalue and Frobenius norm of the matrices $P_{BC}^{(2)}$ and $R_{BC}^{(2)}$ are given by
\begin{eqnarray}
&&\lambda_{min}(P_{BC}^{(2)})= , \lambda_{min}(R_{BC}^{(2)})= 0 \nonumber\\&&
\|P_{BC}^{(2)}\|_{2}^{2}=\frac{7}{25}, \|R_{BC}^{(2)}\|_{2}^{2}=\frac{4}{25}
\label{3ss4}
\end{eqnarray}
Using the information given in (\ref{3ss2}), (\ref{3ss3}) and (\ref{3ss4}), the R.H.S of the inequality (\ref{th4}) gives
\begin{eqnarray}
&&\sum_{i=1}^{2} p_{i}[C_{l_{1}}(\varrho^{(i)}_{A})+2 [(\|P_{BC}^{(i)}\|_{2}^{2}+\|R_{BC}^{(i)}\|_{2}^{2}-\sum^{4}_{j=1} |(\varrho^{(i)}_{BC})_{jj}|^2)^\frac{1}{2} \nonumber\\&+& \sqrt{\lambda_{min}(P_{BC}^{(i)}) \lambda_{min}(R_{BC}^{(i)})}]\times (1+C_{l_{1}}(\varrho^{(i)}_{A}))]\nonumber\\&=& \frac{10+14\sqrt{2}}{25}+\frac{28-14\sqrt{2}}{25}p
\label{3ssrhs}
\end{eqnarray}
One can easily find that $C_{l_{1}}(\xi_{ABC}^{(2)})$ given in (\ref{3sslhs}) is always greater than the quantity given in (\ref{3ssrhs}) for any $p\in [0,1]$. Thus the inequality (\ref{th4}) is violated by the state $\xi_{ABC}^{(2)}$ and hence the tripartite state described by the density operator $\xi_{ABC}^{(2)}$ is entangled.\\
\textbf{Illustration-3:} Let us consider a three-qubit state $\xi_{ABC}^{(3)}$ given by
\begin{eqnarray}
\xi_{ABC}^{(3)} &=& p(|0\rangle\langle 0|+|0\rangle\langle 1|+|1\rangle\langle 0|) \otimes \frac{1}{2}(|00 \rangle\langle 00|+|11 \rangle\langle 11|) \nonumber\\&+& (1-p)|1\rangle\langle 1| \otimes \frac{1}{2}(|01 \rangle\langle 01|+|10 \rangle\langle 10|),~~0\leq p \leq 1
\label{3sss1}
\end{eqnarray}
It can be easily shown that the three-qubit state $\xi_{ABC}^{(3)}$ is a separable state.\\
To verify it via our criterion (\ref{th4}) stated in Theorem-4, we calculate $l_{1}$ norm of coherence of the state $\xi_{ABC}^{(3)}$, which is given by
\begin{eqnarray}
C_{l_{1}}(\xi_{ABC}^{(3)}) = 2p
\label{3sss2}
\end{eqnarray}
We calculate the following quantities required for R.H.S of the inequality (\ref{th4}) with respect to the state $\xi_{ABC}^{(3)}$ and they are given by
\begin{eqnarray}
&&C_{l_{1}}(\rho_{A}^{(1)})=2, C_{l_{1}}(\rho_{A}^{(2)})=0\nonumber\\&&
\|P_{BC}^{(1)}\|_{2}^{2}=\frac{1}{4}, \|R_{BC}^{(1)}\|_{2}^{2}=\frac{1}{4};\nonumber\\&&
\lambda_{min}(P_{BC}^{(1)})=\lambda_{min}(R_{BC}^{(1)})=0;\nonumber\\&&
\|P_{BC}^{(2)}\|_{2}^{2}=\frac{1}{4}, \|R_{BC}^{(2)}\|_{2}^{2}=\frac{1}{4};\nonumber\\&&
\lambda_{min}(P_{BC}^{(2)})=\lambda_{min}(R_{BC}^{(2)})=0
\label{3sss3}
\end{eqnarray}
Now, we are in a position to calculate the R.H.S of the inequality (\ref{th4}) and found out to be $2p$ while the L.H.S of the inequality (\ref{th4}) also found out to be $2p$. The equality condition of (\ref{th4}) is achieved in this case. Thus the inequality (\ref{th4}) is satisfied by $\xi_{ABC}^{(3)}$ and it should be since $\xi_{ABC}^{(3)}$ is a separable state.
\section{Conclusion}
To summarize, we have explored the connection between the coherence and entanglement in the context of detection of entanglement. To established the connection, we use $l_{1}$ norm of coherence of the qubit-qudit system and then derive the lower bound of it using the entanglement criterion. Therefore, the derived lower bound help in the detection of entangled bipartite qubit-qudit states. Also we have shown that if the qubit-qudit syatem is separable then we can achieve the upper bound of the $l_{1}$ norm of coherence of the given separable system. Further, we find that if $l_{1}$ norm of coherence of any bipartite qubit-qudit state is greater than the upper bound then the given qubit-qudit state is entangled. This fact is stated in corollary-1. Moreover, we have applied the entanglement criterion given in corollary-1 to study the entanglement property of tripartite $qubit-qudit-qudit$, $qudit-qubit-qudit$ and $qudit-qudit-qubit$ system. We have illustrated the fact given in corollary-1 and corollary-2 with few examples. We may also note here that corollary-2 may detect entangled tripartite system but it does not take part in the classification of tripartite entangled state. That is, corollary-2 only provide us the information that whether the state under investigation is entangled but it unable to say anything about the class in which the detected entangled state is belonging. Now it remain open for the generalisation of the result obtained in this work to qudit-qudit system.

\section{Data Availability Statement}
Data sharing not applicable to this article as no datasets were generated or analysed during the current study.

\end{document}